# Motivation and Contribution in Open Source Software: A Meta-Analytic Investigation

Parisa Omidmand[1], Saeed Nosratabadi[2]

[1] Department of Economics, Texas Tech University

[2] Rawls College of Business, Texas Tech University

**Abstract**

Research on why people contribute to open source software (OSS) and related open collaboration platforms spans more than two decades, but its findings conflict, and studies often treat the work people contribute and their intention to keep contributing as interchangeable. Drawing on self-determination theory and motivation crowding theory, we meta-analyze 41 studies. The main analysis pools 178 correlations of eight antecedents with contribution effort and continuance intention from 29 independent samples (N = 9,632). Seven antecedents relate positively to effort, all eight relate positively to intention, and each relates at least as strongly to intention as to effort. Contrary to a common assumption, intrinsic motivation is not more strongly related to effort than extrinsic motivation, but its gap between intention and effort is larger. Reputation relates to effort mainly on platforms with a reputation system, and we find no sign that monetary rewards crowd out enjoyment. The findings inform how platforms can turn contributors' intentions into work.

Parisa Omidmand: parisa.omidmand@ttu.edu
Saeed Nosratabadi: snosrata@ttu.edu

# Introduction

Much of today's software and online knowledge is produced by people who are not employed to produce it. The Linux kernel, the Apache web server, and Wikipedia grew largely through voluntary contributions, and similar arrangements now support crowdsourcing contests and citizen science projects (Benkler, 2002; Levine & Prietula, 2014). Firms build products on this work and pay some developers to take part in it (Fitzgerald, 2006). Platform owners try to raise contribution through design, with reputation systems that award points, ranks, and badges, and with monetary rewards such as salaries, prizes, and bounties (Resnick et al., 2000). Contribution is thus a core problem for digital platforms: it keeps them alive, the levers used to raise it are features of the platforms themselves, and most contributors can leave at any time.

Research on why people contribute spans more than two decades (Hars & Ou, 2002; Hertel et al., 2003; Lerner & Tirole, 2002; von Krogh et al., 2012). It has converged on a common set of motivations: enjoyment-based intrinsic motivation, community identification, altruism, reciprocity and own use, reputation, career and human capital, and monetary rewards, along with contributors' belief in their own ability (Lakhani & Wolf, 2005; Wasko & Faraj, 2005). The field agrees on what the motivations are but not on how strongly each one matters, for three reasons.

First, findings conflict. Lakhani and Wolf (2005) found that the creativity developers felt in their work was the strongest driver of their effort, whereas Roberts et al. (2006) found that intrinsic motivation did not raise the contribution levels of Apache developers, while pay and status did. For reciprocity and own use, results range from positive (Hertel et al., 2003) to negative (Roberts et al., 2006) to null (Wasko & Faraj, 2005). Second, studies that appear to examine the same relationship measure different things. Some measure contribution effort, the work people actually do, and others measure continuance intention, the stated plan to keep contributing (Wu et al., 2007). Because intentions predict behavior only in part (Webb & Sheeran, 2006), a motivation can predict one outcome well and the other poorly. Reputation is sometimes a self-reported motive and sometimes a platform score that past contribution produces (Goes et al., 2016), and monetary rewards are sometimes actual pay and sometimes a stated desire for money (Ke & Zhang, 2009). Third, two claims that shape advice to platform owners have not been tested across studies: that intrinsic motivation sustains behavior better than extrinsic motivation (Ryan & Deci, 2000b), and that monetary rewards crowd out intrinsic motivation (Frey &

Jegen, 2001). A platform that adds a reputation system or pays its contributors bets on particular motivations, and a wrong bet can waste money or weaken volunteer contribution.

We therefore conduct a meta-analysis that pools the evidence while keeping outcomes and measures apart. It makes four contributions. First, unlike narrative reviews (Crowston et al., 2012; von Krogh et al., 2012), it pools the correlations of eight antecedents with contribution effort and with continuance intention, centering on OSS and using related settings as a comparison. Second, it tests the ordering of motivations proposed by self-determination theory separately for effort and intention; earlier meta-analyses examined performance in school and work (Cerasoli et al., 2014) and knowledge sharing in organizations (Nguyen et al., 2019), not voluntary contribution to shared products. Third, it links a platform design feature, the reputation system, to the strength of a motivation. Fourth, it tests three predictions of motivation crowding theory. We find that motivation explains what contributors intend better than what they do, that intrinsic motivation is not more strongly related to effort than extrinsic motivation but has a larger gap between intention and effort, that reputation relates to effort mainly where a reputation system exists, and that money does not crowd out enjoyment.

# Theoretical Background and Hypotheses

Open collaboration is production in which loosely coordinated participants create goods of economic value and make them available to contributors and noncontributors alike (Levine & Prietula, 2014). OSS is its most studied form; wikis, crowdsourcing contests (Boudreau et al., 2011), citizen science projects (Nov et al., 2014), and online communities (Wasko & Faraj, 2005) are related forms. Much of the research on why people contribute builds on self-determination theory, which defines intrinsic motivation as doing an activity for its inherent satisfaction and extrinsic motivation as doing it to attain a separable outcome (Deci & Ryan, 1985; Ryan & Deci, 2000a); extrinsic motivation varies in how far its goal is internalized (Gagné & Deci, 2005). Economic accounts add that visible work signals talent (Lerner & Tirole, 2002). Building on three classifications of motivations (Hars & Ou, 2002; Lakhani & Wolf, 2005; von Krogh et al., 2012), we examine eight antecedents: three in an intrinsic group, three in an extrinsic group, and two in neither group (Figure 1).

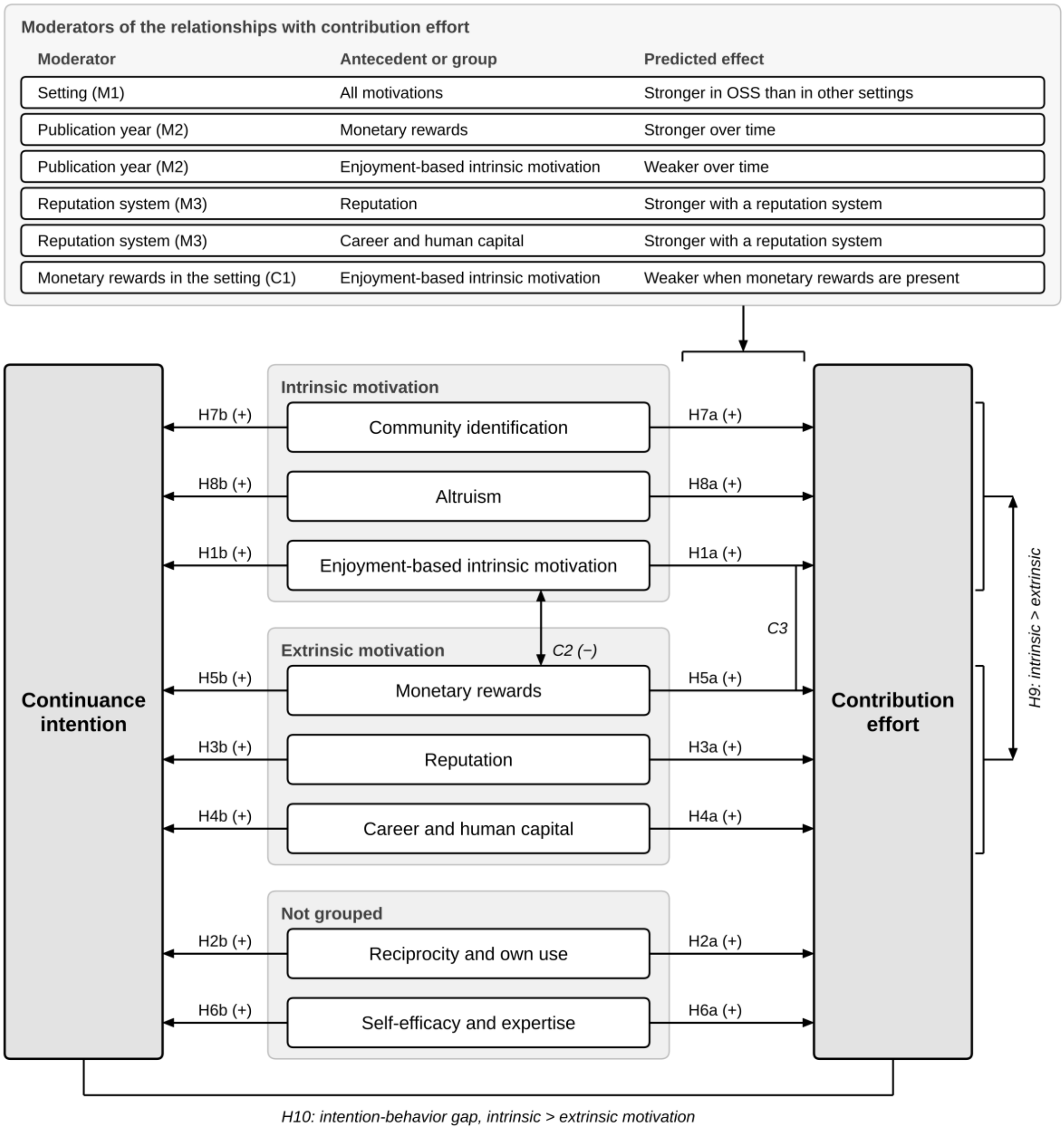

**Figure 1: Research framework**

*Note.* H1a to H8a link each antecedent to contribution effort, and H1b to H8b link each antecedent to continuance intention; (+) and (−) show the predicted direction. H9 and H10 compare the intrinsic and extrinsic groups on effort and on the intention-behavior gap; > shows the predicted ordering. M1 to M3 are moderator hypotheses, and C1 to C3 are motivation crowding hypotheses.

# Construct Definitions and Measurement

Table 1 defines the constructs and names the theory behind each hypothesis. We assign measures to constructs by item content rather than by label, because studies use the same label for different content and different labels for the same content. Most studies measure reputation as a motive, the importance placed on recognition from peers (Hars & Ou, 2002), but some use indicators that the platform displays, such as follower counts or points (Restivo & van de Rijt, 2014). Because these indicators accrue from past contribution, we use them only in a robustness check. Monetary rewards are measured as actual payment (Roberts et al., 2006) or as an expected or desired monetary return (Ke & Zhang, 2009); we pool both and compare them in a robustness check. For self-efficacy and expertise (Kankanhalli et al., 2005), we code general development experience but not tenure or past activity, which partly record past contribution. Regulatory styles (Ke & Zhang, 2010; Li et al., 2012) are assigned by item content: external regulation about money to monetary rewards, identified regulation to community identification or to career and human capital, and introjected regulation to reputation only when its items describe recognition from others. Contribution effort includes hours, self-rated effort, and contributions recorded in digital trace data, such as commits and edits. Continuance intention is the stated intention to keep contributing (Bhattacherjee, 2001); we reverse turnover intention. We do not examine ideology (Stewart & Gosain, 2006), because few studies report it and they define it differently.

| Construct | Group | Definition | Labels coded under the construct | Theoretical basis (hypothesis) |
|---|---|---|---|---|
| Enjoyment-based intrinsic motivation | Intrinsic | Contributing because the work itself is enjoyable or stimulating | Fun, enjoyment, flow, intellectual stimulation, creativity; broad intrinsic scales dominated by fun | Self-determination theory (Ryan & Deci, 2000b); H1 |
| Reciprocity and own use | Not grouped | Contributing because one needs the product, or to return help received | Own use, use value, personal need, pragmatic motives, reciprocity, obligation to give back as a user | Private-collective innovation model; norm of reciprocity (Gouldner, 1960; von Hippel & von Krogh, 2003); H2 |
| Reputation | Extrinsic | Contributing to gain recognition and standing among peers | Reputation, peer recognition, status, performance orientation, competitive motives | Signaling theory (Lerner & Tirole, 2002; Spence, 1973); H3 |
| Career and human capital | Extrinsic | Contributing to build skills or improve job prospects | Career advancement, self-marketing, human capital, learning, skill improvement | Human capital theory (Becker, 1964); H4 |
| Monetary rewards | Extrinsic | Being paid for contributing, or expecting money from it | Paid status, paid hours, prizes, bounties, revenue from related products, desire for money, external regulation | Pay as a reason and the time to contribute (Roberts et al., 2006); H5 |
| Self-efficacy and expertise | Not grouped | Belief in one's ability to contribute, and the skill itself | Self-efficacy, perceived competence, self-rated expertise, general development experience | Social cognitive theory (Bandura, 1997); H6 |
| Community identification | Intrinsic | Seeing oneself as a member of the community | Identification, affective commitment, sense of community, obligation-based intrinsic motivation | Social identity theory (Tajfel & Turner, 1979); H7 |
| Altruism | Intrinsic | Contributing to help others or to advance the shared goals of the project | Altruism, helping others, collective motives, prosocial values | Functional approach to volunteering (Clary et al., 1998); H8 |
| Contribution effort | Outcome | Work actually contributed | Hours per week, effort scales, commits, posts, edits, reports, classifications; self-reported or from digital trace data | Outcome of H1a to H8a |
| Continuance intention | Outcome | Stated intention to keep contributing | Continuance intention, intention to contribute, participation intention, intention to remain, turnover intention (reversed) | Outcome of H1b to H8b |

**Table 1: Construct definitions, coded measures, and theoretical basis**

*Note.* Measures are assigned by item content, not by label. Reputation indicators displayed by the platform enter only a robustness check. Tenure, account age, and past activity are not coded as expertise.

# Motivations, Contribution Effort, and Continuance Intention

We expect each antecedent to relate positively to both outcomes, and each pair of hypotheses rests on an established theory (Table 1). Enjoyable work is its own reward (Ryan & Deci, 2000b). People contribute to software they need (von Hippel & von Krogh, 2003) and return help received (Gouldner, 1960). Visible contributions signal ability to peers and employers (Spence, 1973), and the skills built pay off in the labor market (Becker, 1964; Hann et al., 2013). Pay gives people a reason and the time to contribute (Roberts et al., 2006). People who believe they can perform a task invest more effort (Bandura, 1997), people work for groups that are part of their self-concept (Tajfel & Turner, 1979), and helping others is a core reason to volunteer (Clary et al., 1998). Thus, we propose:

> **H1a to H8a:** Enjoyment-based intrinsic motivation (H1), reciprocity and own use (H2), reputation (H3), career and human capital (H4), monetary rewards (H5), self-efficacy and expertise (H6), community identification (H7), and altruism (H8) are each positively related to contribution effort.

> **H1b to H8b:** Each of these antecedents is positively related to continuance intention.

## Intrinsic versus Extrinsic Motivation

Self-determination theory also orders motivations. Intrinsically motivated behavior is its own reward, whereas behavior driven by an external payoff lasts only while the payoff remains attractive and within reach (Ryan & Deci, 2000b). In open collaboration, a reputation or a better job may come years later, if at all (Lerner & Tirole, 2002), whereas enjoyment, belonging, and helping pay off with each contribution. The OSS literature often assumes this ordering (Shah, 2006), although incentives predict the quantity of performance better than its quality (Cerasoli et al., 2014). Because most contributors are volunteers, we expect the ordering to hold for effort. Following Hars and Ou (2002) and von Krogh et al. (2012), the intrinsic group covers enjoyment-based intrinsic motivation, community identification, and altruism, and the extrinsic group covers reputation, career and human capital, and monetary rewards, which serve a separate outcome (Ryan & Deci, 2000a). We leave reciprocity and own use out of the main test, because the classifications disagree on it (Lakhani & Wolf, 2005), and add it to the extrinsic group in a robustness check. Self-efficacy and expertise is a belief about ability rather than a motivation. Thus, we propose:

> **H9:** Intrinsic motivation is more strongly related to contribution effort than extrinsic motivation.

## The Intention-Behavior Gap

Intentions turn into behavior only in part (Ajzen, 1991; Webb & Sheeran, 2006), and in IS research, habit limits how well intention predicts continued use (Limayem et al., 2007). We call the difference between a motivation's relationship with continuance intention and with contribution effort the intention-behavior gap and expect it to be larger for intrinsic motivation, for two reasons. First, intentions are carried out only when people control the resources they require, such as time (Ajzen, 1991). A firm that pays a developer also sets aside working hours, much OSS work is done during regular working hours (Riehle et al., 2014), and paid contributors spend more hours than unpaid ones (Lakhani & Wolf, 2005). A volunteer must find the time outside work. Second, enjoyment, belonging, and helping are socially desirable reasons to contribute, so respondents may overstate them in line with their plans to continue, whereas being paid is a fact. General method variance inflates all relationships measured in the same survey (Podsakoff et al., 2003), so by itself it cannot produce a larger gap for one group. Thus, we propose:

**H10:** The intention-behavior gap is larger for intrinsic motivation than for extrinsic motivation.

## Moderators

**Setting.** An OSS contribution, such as a code patch, often takes hours of work and enough skill to pass review (von Krogh et al., 2003), whereas a wiki edit or a citizen science classification can take minutes. Fine-grained tasks let people with little motivation contribute (Benkler, 2002), so motivation should matter less where contributing is cheap. Because our search was designed for OSS, this test is exploratory.

**M1:** The relationships between the motivations and contribution effort are stronger in OSS than in other open collaboration settings.

**Publication year.** OSS has moved from a volunteer movement toward a commercial form in which firms pay developers (Fitzgerald, 2006; Riehle et al., 2014), so pay should become a stronger driver of effort over time and enjoyment a weaker one. We measure time as the year a sample was first reported.

**M2:** Over time, the relationship between monetary rewards and contribution effort becomes stronger, and the relationship between enjoyment-based intrinsic motivation and contribution effort becomes weaker.

**Reputation system.** A signal works only if others can observe it (Spence, 1973), and the incentive to signal talent grows with the visibility of performance (Lerner & Tirole, 2002). A reputation system makes each contribution visible as points, ranks, or badges (Resnick et al., 2000). Users of an online knowledge exchange contributed more as they approached the next status level (Goes et al., 2016), and symbolic awards raised the retention of new Wikipedia editors (Gallus, 2017). We code a reputation system as present when the platform displays public, quantified, or formal status markers, as on GitHub, Stack Exchange, and Kaggle and in the Apache projects.

**M3:** The relationships of reputation and of career and human capital with contribution effort are stronger on platforms with a reputation system than on platforms without one.

## Motivation Crowding

Platforms and firms increasingly bring money into settings that once relied on volunteers, through prizes (Boudreau et al., 2011) and salaries (Fitzgerald, 2006). Motivation crowding theory holds that rewards crowd out intrinsic motivation when people perceive them as controlling and crowd it in when they perceive them as supportive (Frey & Jegen, 2001). Expected, performance-contingent rewards tend to undermine intrinsic motivation (Deci et al., 1999), and payment has reduced volunteers' effort in the field (Gneezy & Rustichini, 2000), although pay did not undermine the motivation of Apache developers (Roberts et al., 2006). The theory implies three testable predictions. If money displaces enjoyment, enjoyment should matter less where some contributors are paid or prizes are offered; people who are paid or value money should report less enjoyment; and because perceptions of money differ across settings, its relationship with effort should vary more across studies than that of enjoyment. Thus, we propose:

> **C1:** The relationship between enjoyment-based intrinsic motivation and contribution effort is weaker in settings where monetary rewards are present than where they are absent.
>
> **C2:** Enjoyment-based intrinsic motivation and monetary rewards are negatively related.
>
> **C3:** The relationship between monetary rewards and contribution effort is more heterogeneous across studies than the relationship between enjoyment-based intrinsic motivation and contribution effort.

# Method

## Literature Search

We followed the PRISMA 2020 guidelines (Page et al., 2021). On 9 June 2026, two authors searched titles, abstracts, and author keywords in Web of Science Core Collection (7,048 records), Scopus (9,230), ACM Digital Library (2,179), Business Source Complete (580), PsycINFO (271), EconLit (102), and AIS eLibrary (90); IEEE Xplore returned no records. To reduce publication bias, we also searched ProQuest Dissertations & Theses (1,090) and SSRN (369) (Rosenthal, 1979). The search combined three blocks of terms with AND, with 43 terms in total (Table 2), and was limited to English but not by date or document type. A pilot on 1,000 Web of Science records led us to drop "OSS" and "learning." The final search retrieved 18 of 20 known relevant studies; the two it missed are IEEE publications. It returned 20,959 records, and after removing 7,122 duplicates and one non-English record, 13,836 remained (Figure 2). Studies of related open collaboration settings retrieved by the search form a comparison set.

| **Block** | **Terms** |
|---|---|
| Setting (9 terms) | "open source" OR "open-source" OR FLOSS OR FOSS OR "free software" OR "peer production" OR GitHub OR SourceForge OR "software commons" |
| Antecedents (20 terms) | motivation OR motive* OR intrinsic OR extrinsic OR "self-determination" OR reputation OR recognition OR status OR signal* OR "career concern*" OR "human capital" OR incentive* OR paid OR sponsor* OR compensat* OR volunteer OR ideology OR identification OR reciprocity OR "self-efficacy" |
| Outcomes (14 terms) | contribut* OR participat* OR effort OR commit* OR "lines of code" OR "pull request*" OR retention OR turnover OR continuance OR "sustained participation" OR persistence OR disengage* OR abandon* OR dropout |

**Table 2: Search blocks**

*Note.* The three blocks were combined with AND. Each block was searched in titles, abstracts, and author keywords; in Web of Science, the topic field also covers Keywords Plus.

## Screening, Eligibility, and Coding

Two authors independently screened a random sample of 300 titles and abstracts and agreed on 297 (99%; Cohen's κ = .87; Cohen, 1960), which is almost perfect agreement (Landis & Koch, 1977), and then divided the remaining records; 603 were retained. At the full-text stage, they agreed on 117 of a random sample of 120 articles (97.5%; κ = .89). Disagreements were resolved by discussion. Of the 603 articles, 553 were excluded and 50 met the criteria; nine were additional reports of included samples and were merged, which left 41 studies. A study was included if it reported original quantitative data on individual contributors to OSS or another open collaboration setting, measured at least one antecedent as a characteristic of the contributor and as a predictor of contribution effort or continuance intention, reported a zero-order correlation or a convertible statistic (or, for the expanded analysis, regression results), and was written in English. Online Appendix A gives examples of excluded reports. Each independent sample in an article, such as the three citizen science projects in Nov et al. (2014), was treated separately.

The main analysis uses 178 zero-order correlations from 29 independent samples in 27 studies (N = 9,632), first reported between 2002 and 2026. Of these samples, 24 come from journal articles or conference papers and 5 from dissertations; 18 come from OSS and 11 from other settings (5 crowdsourcing contests, 3 citizen science projects, 2 online communities, and 1 open government platform). Eighteen use surveys alone, 10 combine surveys with platform records, and 1 uses platform records only. Effort is measured in 20 samples, 9 of them with digital trace data, and intention in 15; 6 measure both. The expanded analysis adds partial correlations and covers 37 samples in 35 studies (N = 17,630). Three further studies contribute only to C2, and three studies with six samples contribute only platform reputation indicators. Appendices B and C list the studies, their coding, and their correlations.

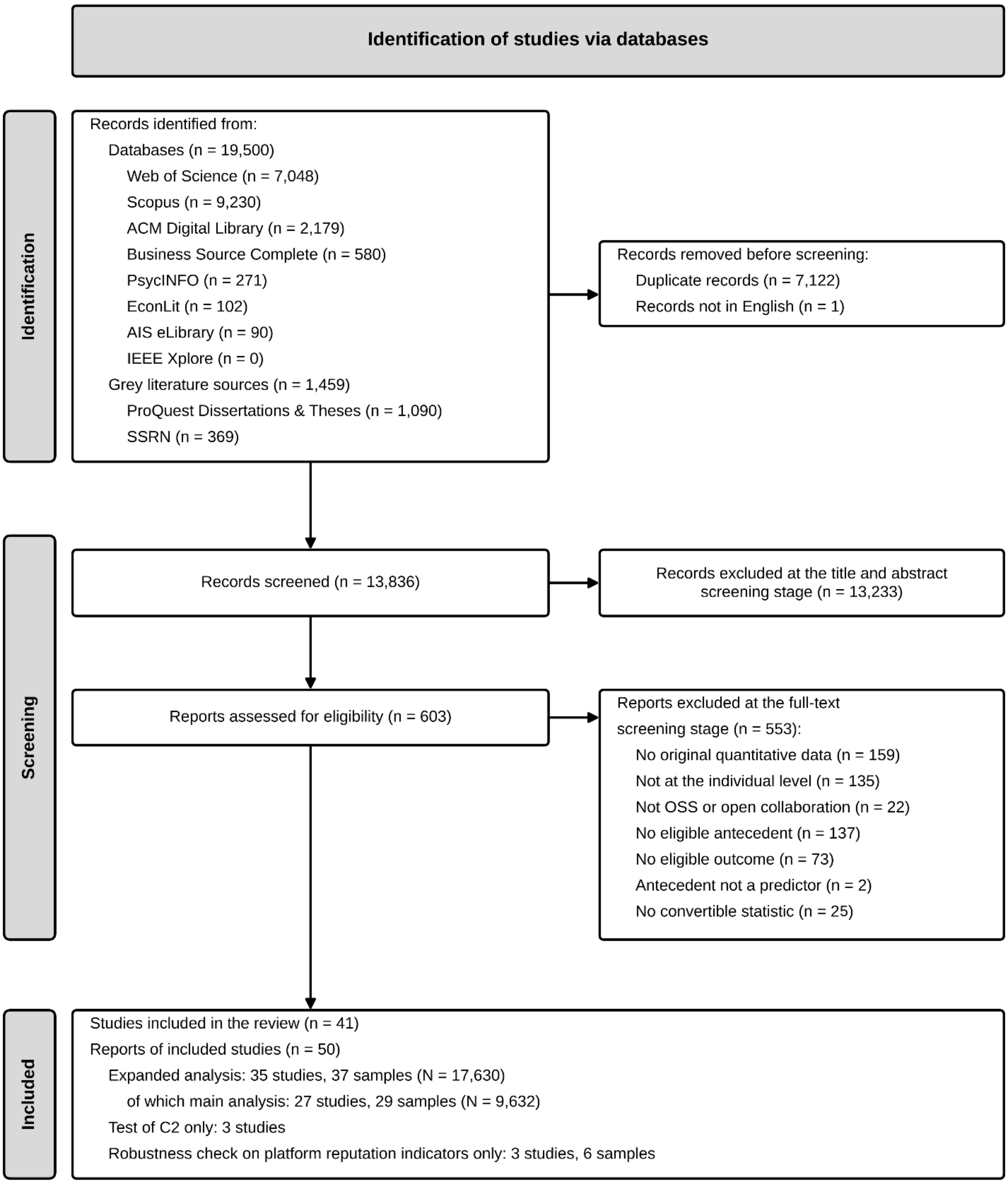

**Figure 2: PRISMA 2020 flow diagram**

*Note.* Adapted from Page et al. (2021). n = number of records, reports, or studies; N = total sample size.

Two authors coded each effect (antecedent, outcome, statistic, sample size, effect type, and whether effort was self-reported or recorded in digital trace data) and each sample (setting, year of the earliest report, design, presence of monetary rewards, presence of a reputation system, and publication status). Both coders independently coded all 41 studies and identified 236 of the 247 effects (95.5%); the other 11 were added after discussion. Agreement was κ = .96 for the outcome, .93 for the antecedent, .94 for the effect type, .95 for the outcome source, and .89 for the type of reward. Correlations agreed within .01 for 97.0% of effects, and sample sizes were identical for 97.9%. For the 46 samples, κ was .96 for setting, .82 for monetary rewards, .76 for reputation system, and 1.00 for publication status, which is substantial or almost perfect agreement (Landis & Koch, 1977).

# Analysis

We pooled correlations with random-effects models (Borenstein et al., 2009). We converted each correlation to Fisher's z, weighted it by the inverse of $1/(N - 3)$ plus the between-study variance $\tau^2$, estimated with the DerSimonian and Laird (1986) method, and transformed pooled estimates back to correlations. We did not correct for measurement error, so our estimates are conservative (Schmidt & Hunter, 2015). We computed composite correlations from item-level correlations (Schmidt & Hunter, 2015) and converted other statistics with standard formulas (Lipsey & Wilson, 2001). The main analysis uses only zero-order correlations. The expanded analysis, a robustness check, adds partial correlations, $t/\sqrt{(t^2 + df)}$, for samples that report only regression results (Stanley & Doucouliagos, 2012). Each sample contributes one effect per relationship; we averaged multiple effects and used their largest N, capped at 10,000. We pooled a relationship only when at least three samples reported it, and we report 95% confidence intervals, Q, $I^2$ (Higgins et al., 2003), and 80% credibility intervals.

H1 to H8 are supported when the pooled correlation is positive and significant. For H9, we tested the difference between the groups' pooled Fisher's z with effort and, because many samples contribute to both groups, also pooled within-sample differences, assuming a correlation of 0, .3, or .5 between group means. For H10, each group's gap is its pooled Fisher's z with intention minus its pooled Fisher's z with effort; we tested the difference between the gaps with a z test. We tested M1, M3, and C1 with subgroup analyses that required at least two samples per subgroup and excluded samples coded as unclear, and M2 with a meta-regression on year, centered at 2015 and scaled per decade, with the Knapp-Hartung adjustment (Knapp & Hartung, 2003). For C2, we pooled the within-sample correlations between enjoyment and monetary rewards, and for C3, we compared $\tau^2$ with Q-profile confidence intervals.

Because common method variance can inflate self-reported relationships (Podsakoff et al., 2003), we repeated the main tests with self-reported and trace-based effort separately, with published studies only, and with surveys only; added reciprocity and own use to the extrinsic group for H9; and compared the reputation motive with platform reputation indicators and actual payment with expected money. We reran H9 and H10 after dropping each sample in turn (Viechtbauer & Cheung, 2010) and reran the moderator tests without one sample with very high correlations and without the three citizen science samples from one article. For publication bias, we compared published with unpublished studies and, for the seven relationships with at least 10 samples, applied Egger's test (Egger et al., 1997) and trim-and-fill (Duval & Tweedie, 2000). We used R 4.5.0 and metafor 5.0-1. The search records, coded data, and analysis script are available at Mendeley Data (DOI: 10.17632/kdkxpgtgr9.1).

# Results

## Motivations, Contribution Effort, and Continuance Intention

All eight antecedents were positively related to continuance intention, and all except reciprocity and own use were positively related to contribution effort (Table 3). Reciprocity and own use correlated $r = .16$ with effort, but its confidence interval included zero. The relationship between monetary rewards and intention rests on only three samples and was not significant in the expanded analysis ($r = .18$, 95% CI [−.02, .37], $k = 5$); no other conclusion changed (Appendix D). Correlations with effort ranged from $r = .11$ (career and human capital) to $r = .28$ (community identification), and correlations with intention were larger, from $r = .22$ to $r = .53$ (community identification and altruism). Every antecedent was at least as strongly related to intention as to effort; the difference was largest for altruism and community identification and smallest for monetary rewards ($r = .25$ for both). $I^2$ exceeded 80% for 14 of the 16 relationships, and the 80% credibility intervals included zero for five relationships with effort and three with intention.

| Hypothesis | Antecedent | Outcome | k | N | r | 95% CI | 80% CV | Q | I² (%) | Supported |
|---|---|---|---|---|---|---|---|---|---|---|
| H1a | Enjoyment-based intrinsic motivation | Effort | 14 | 6,728 | .24 | [.12, .36] | [−.05, .50] | 322.56*** | 96 | Yes |
| H1b | Enjoyment-based intrinsic motivation | Intention | 9 | 2,994 | .40 | [.31, .48] | [.24, .54] | 54.44*** | 85 | Yes |
| H2a | Reciprocity and own use | Effort | 8 | 1,753 | .16 | [−.10, .41] | [−.31, .57] | 212.16*** | 97 | No |
| H2b | Reciprocity and own use | Intention | 7 | 1,253 | .25 | [.02, .45] | [−.14, .57] | 105.05*** | 94 | Yes |
| H3a | Reputation | Effort | 11 | 5,077 | .26 | [.09, .42] | [−.11, .57] | 348.10*** | 97 | Yes |
| H3b | Reputation | Intention | 6 | 1,244 | .36 | [.19, .52] | [.09, .59] | 56.34*** | 91 | Yes |
| H4a | Career and human capital | Effort | 4 | 547 | .11 | [.02, .20] | [.07, .15] | 3.40 | 12 | Yes |
| H4b | Career and human capital | Intention | 7 | 1,170 | .22 | [.09, .34] | [.01, .41] | 32.09*** | 81 | Yes |
| H5a | Monetary rewards | Effort | 10 | 2,186 | .25 | [.14, .35] | [.04, .43] | 59.36*** | 85 | Yes |
| H5b | Monetary rewards | Intention | 3 | 630 | .25 | [.01, .47] | [−.00, .48] | 19.37*** | 90 | Yes[a] |
| H6a | Self-efficacy and expertise | Effort | 8 | 4,258 | .25 | [.07, .42] | [−.08, .53] | 212.19*** | 97 | Yes |
| H6b | Self-efficacy and expertise | Intention | 5 | 862 | .32 | [.05, .54] | [−.06, .61] | 61.09*** | 93 | Yes |
| H7a | Community identification | Effort | 8 | 1,348 | .28 | [.04, .49] | [−.15, .63] | 144.32*** | 95 | Yes |
| H7b | Community identification | Intention | 7 | 1,505 | .53 | [.34, .68] | [.19, .76] | 123.78*** | 95 | Yes |
| H8a | Altruism | Effort | 6 | 3,675 | .21 | [.15, .26] | [.15, .26] | 9.92 | 50 | Yes |
| H8b | Altruism | Intention | 7 | 1,216 | .53 | [.29, .70] | [.10, .79] | 148.98*** | 96 | Yes |

**Table 3: Relationships of the eight antecedents with contribution effort and continuance intention**

*Note.* Main analysis with zero-order correlations only. k = number of independent samples; N = total sample size; r = random-effects pooled correlation; CI = confidence interval; CV = credibility interval; Q = test of homogeneity; $I^2$ = percentage of observed variance due to true differences between studies. Supported = r positive and significant ($p < .05$). [a] Based on three samples; not significant in the expanded analysis. *** $p < .001$.

# Intrinsic versus Extrinsic Motivation and the Intention-Behavior Gap

Intrinsic motivation was not more strongly related to effort than extrinsic motivation (Table 4). The intrinsic group correlated r = .21 with effort and the extrinsic group r = .28 ($\Delta z = -.07$, 95% CI [−.24, .09], p = .369). In the 14 samples that report both groups, the within-sample difference was similar ($\Delta z = -.06$, p = .205; p = .204 and .198 with assumed correlations of .3 and .5). H9 was not supported.

The intention-behavior gap was larger for intrinsic motivation. The intrinsic group correlated r = .49 with intention and r = .21 with effort (gap $\Delta z = .32$), whereas the extrinsic group correlated r = .32 and r = .28 (gap $\Delta z = .04$). The difference between the gaps was $\Delta z = .28$ (95% CI [.02, .54], p = .033) and $\Delta z = .30$ (p = .017) in the expanded analysis. Only three samples report both outcomes for both groups; in two, the gap was larger for intrinsic motivation ($\Delta z = .12$ and .25), and in the third, the gaps were nearly equal ($\Delta z = -.02$). The result also depends partly on one flagged value, a correlation of .89 between altruism and intention in one dissertation; without it, $\Delta z = .22$ (p = .078). H10 was supported, with these caveats.

*Panel A. Pooled correlations by group*

| Group | Outcome | k | N | r | 95% CI | 80% CV | $I^2$ (%) |
|---|---|---|---|---|---|---|---|
| Intrinsic motivation | Contribution effort | 16 | 7,046 | .21 | [.11, .31] | [−.03, .43] | 94 |
| Extrinsic motivation | Contribution effort | 18 | 6,263 | .28 | [.16, .39] | [−.04, .55] | 95 |
| Intrinsic motivation | Continuance intention | 15 | 4,000 | .49 | [.37, .59] | [.19, .70] | 94 |
| Extrinsic motivation | Continuance intention | 10 | 1,830 | .32 | [.18, .44] | [.05, .54] | 89 |

*Panel B. Contrasts*

| Hypothesis | Contrast | Δz | 95% CI | p | Supported |
|---|---|---|---|---|---|
| H9 | Intrinsic minus extrinsic, effort, across samples | −.07 | [−.24, .09] | .369 | No |
| H9 | Intrinsic minus extrinsic, effort, within 14 samples | −.06 | [−.15, .03] | .205 | No |
| H10 | Intrinsic gap minus extrinsic gap | .28 | [.02, .54] | .033 | Yes |

**Table 4: Tests of H9 and H10**

*Note.* Intrinsic = enjoyment-based intrinsic motivation, community identification, and altruism; extrinsic = reputation, career and human capital, and monetary rewards. Each sample's group score is the mean of its construct-level correlations. Δz = difference in Fisher's z; gap = Fisher's z with intention minus Fisher's z with effort. The within-sample test assumes no correlation between the two group means; assuming .3 or .5 gives p = .204 and p = .198.

# Moderators and Motivation Crowding

Setting did not moderate the group relationships (Table 5). Of five antecedents with at least two samples per setting, only monetary rewards differed (OSS r = .32, other settings r = .05, Δz = .28, $p < .001$; Appendix E), an exploratory result. Publication year did not moderate the relationship of monetary rewards with effort (slope = −.01 per decade, p = .903), and the relationship of enjoyment with effort grew stronger, not weaker (slope = .29, p = .004), although this trend depends on one study from 2023 (without it, slope = .18, p = .078). Reputation was related to effort on platforms with a reputation system (r = .35) but not without one (r = .00; Δz = .37, $p < .001$), and the difference held without the sample with the highest correlations (p = .004) and without the citizen science samples (p = .035). M3 could not be tested for career and human capital.

Enjoyment was more weakly related to effort where monetary rewards were present (r = .10 vs. .51), but three of the four samples without rewards come from one article, so this is not a valid test of C1. Enjoyment and monetary rewards were positively related (r = .21, 95% CI [.07, .34], k = 11), contrary to C2, and the relationship of money with effort varied less across studies than that of enjoyment ($\tau^2$ = .027 vs. .055), contrary to C3. The data thus show no undermining effect.

| Hypothesis | Relationship | Subgroup or test | k | N | r [95% CI] | Δz or slope | p | Supported |
|---|---|---|---|---|---|---|---|---|
| M1 | Intrinsic group → effort | OSS | 10 | 2,999 | .20 [.04, .36] | | | |
| | | Other settings | 6 | 4,047 | .22 [.10, .34] | −.02 | .869 | No |
| M1 | Extrinsic group → effort | OSS | 12 | 2,216 | .31 [.18, .43] | | | |
| | | Other settings | 6 | 4,047 | .23 [−.01, .44] | .09 | .547 | No |
| M1 | Monetary rewards → effort | OSS | 7 | 1,317 | .32 [.23, .41] | | | |
| | | Other settings | 3 | 869 | .05 [−.07, .17] | .28 | < .001 | Yes |
| M2 | Monetary rewards → effort | Slope per decade | 10 | 2,186 | | −.01 | .903 | No |
| M2 | Enjoyment-based intrinsic motivation → effort | Slope per decade | 14 | 6,728 | | .29 | .004 | No (opposite direction) |
| M3 | Reputation → effort | Reputation system present | 8 | 4,233 | .35 [.20, .49] | | | |
| | | Reputation system absent | 3 | 844 | .00 [−.10, .11] | .37 | < .001 | Yes |
| M3 | Career and human capital → effort | Present vs. absent | 3 vs. 1 | | | | | Not testable |
| C1 | Enjoyment-based intrinsic motivation → effort | Rewards present | 9 | 3,125 | .10 [.06, .13] | | | |
| | | Rewards absent | 4 | 3,485 | .51 [.40, .60] | −.46 | < .001 | Not a valid test |
| C2 | Enjoyment-based intrinsic motivation and monetary rewards | Pooled correlation | 11 | 2,560 | .21 [.07, .34] | | .003 | No (positive) |
| C3 | Monetary rewards → effort vs. enjoyment-based intrinsic motivation → effort | $\tau^2$ [95% CI] | 10 vs. 14 | | .027 [.010, .116] vs. .055 [.026, .145] | | | No |

**Table 5: Moderator and motivation crowding tests**

*Note.* Δz = difference in Fisher's z between subgroups (z test). Slope = change in Fisher's z per decade (Knapp-Hartung test). $\tau^2$ = between-study variance with its Q-profile 95% CI. Samples coded as unclear are excluded from M3 and C1. C1 is not a valid test because three of the four samples without rewards come from one article.

# Robustness and Publication Bias

The H9 result held in every robustness check (Appendix D), with differences between Δz = −.11 and −.03 (all $p \geq .250$ in the leave-one-out analysis). The difference between the gaps (H10) stayed positive in every specification (Δz = .22 to .32) but was not significant in published studies only (p = .111), in surveys only (p = .140), or with self-reported effort only (p = .170), and it lost significance when we dropped the sample with the flagged value (p = .091), a sample of episodic OSS volunteers (p = .059), or a sample of question-and-answer contributors (p = .050). With trace-based effort, the intrinsic and extrinsic groups correlated r = .17 and .28 with effort; with self-reported effort, r = .25 and .28. Platform reputation indicators correlated r = .36 with effort (95% CI [−.25, .76]), actual payment r = .34, and expected or desired money r = .17. Restricting the analysis to published studies changed group correlations by .06 or less. Egger's test was significant for reputation (p = .044) and the extrinsic group (p = .018) with effort, but smaller samples reported weaker, not stronger, correlations, and trim-and-fill imputed studies only on the side of larger correlations (Table 6). We found no sign of publication bias.

| Relationship | k | r | Egger intercept (SE) | p | Studies imputed | Adjusted r [95% CI] |
|---|---|---|---|---|---|---|
| Enjoyment-based intrinsic motivation → effort | 14 | .24 | −3.19 (2.74) | .267 | 1 | .26 [.14, .37] |
| Reputation → effort | 11 | .26 | −7.29 (3.11) | .044 | 0 | .26 [.09, .42] |
| Monetary rewards → effort | 10 | .25 | 4.29 (2.58) | .135 | 0 | .25 [.14, .35] |
| Intrinsic group → effort | 16 | .21 | −1.79 (2.10) | .410 | 5 | .29 [.20, .37] |
| Extrinsic group → effort | 18 | .28 | −5.33 (2.03) | .018 | 2 | .32 [.20, .42] |
| Intrinsic group → intention | 15 | .49 | 2.41 (2.92) | .423 | 0 | .49 [.37, .59] |
| Extrinsic group → intention | 10 | .32 | −11.10 (4.86) | .052 | 1 | .35 [.22, .46] |

**Table** 6**: Publication bias tests**

*Note.* Relationships with at least 10 samples in the main analysis. A positive Egger intercept means that smaller samples report larger correlations. Trim-and-fill used the L0 estimator with a random-effects model; all imputed studies were added on the side of larger correlations.

# Discussion

This study offers, to our knowledge, the first quantitative synthesis of why people contribute to OSS and related open collaboration settings; earlier syntheses were narrative (Crowston et al., 2012; von Krogh et al., 2012). Four findings stand out. First, every antecedent relates at least as strongly to continuance intention as to contribution effort. Second, intrinsic motivation is not more strongly related to effort than extrinsic motivation, but its gap between intention and effort is larger. Third, reputation relates to effort mainly on platforms with a reputation system. Fourth, the evidence shows no sign that monetary rewards crowd out enjoyment. Figure 3 updates the research framework with these results.

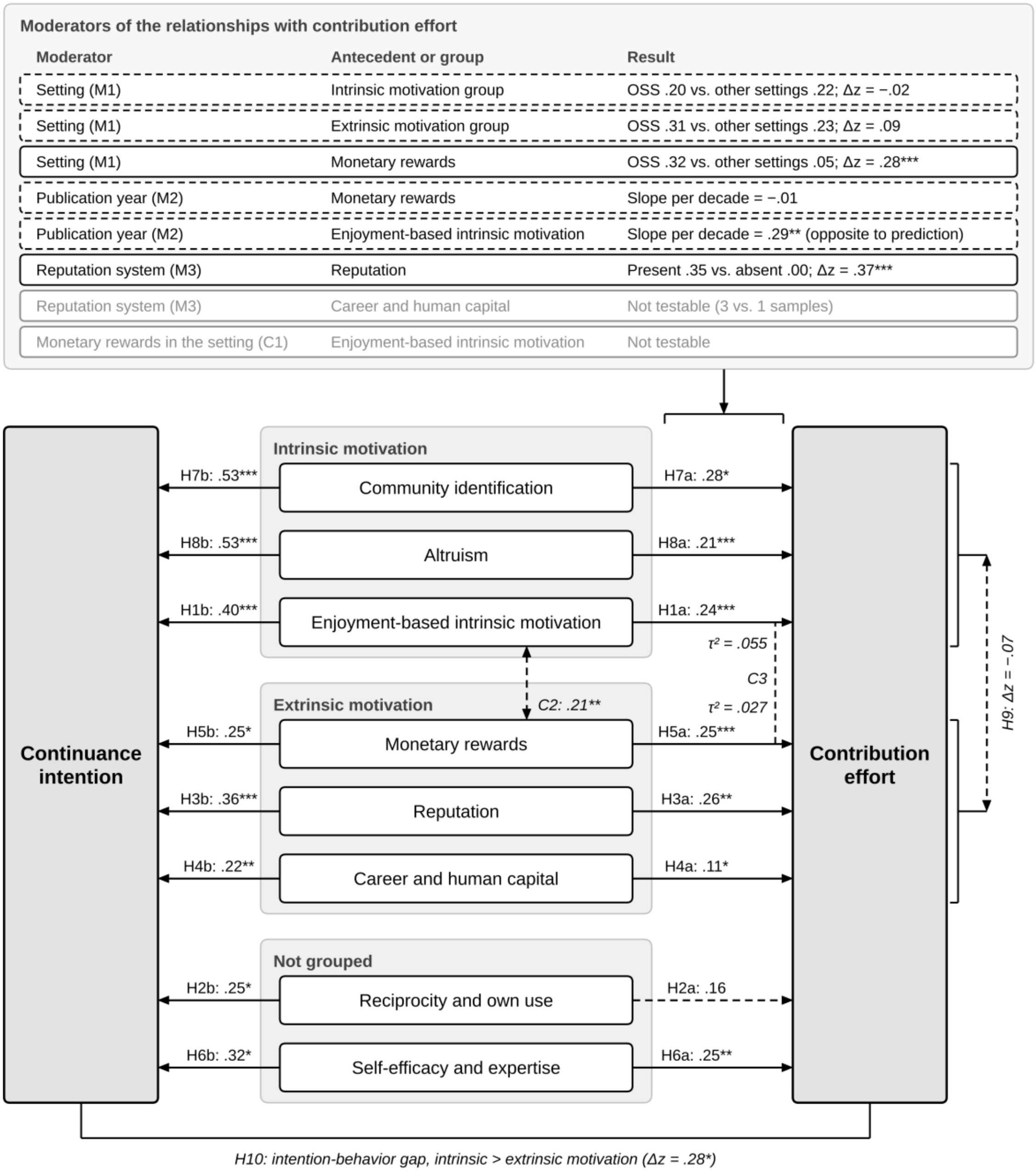

**Figure 3: Updated research framework**

*Note.* Values are pooled correlations from the main analysis unless labeled otherwise. Δz = difference in Fisher's z (intrinsic minus extrinsic for H9 and H10); slope = change in Fisher's z per decade; $\tau^2$ = between-study variance. Solid lines show supported hypotheses, dashed lines show hypotheses that were not supported, and grey elements could not be tested. The setting comparison (M1) is exploratory. * $p < .05$; ** $p < .01$; *** $p < .001$.

# Intrinsic and Extrinsic Motivation

The OSS literature has long treated intrinsic motivation as the more reliable source of contribution. Our results do not support this view for effort. Extrinsic motivation was at least as strongly related to effort in every robustness check, and enjoyment did not lose importance over time. Two explanations seem plausible. First, self-determination theory predicts that autonomous motivation, which includes well-internalized extrinsic goals, sustains behavior about as well as intrinsic motivation does. In OSS, reputation and learning are often internalized goals rather than outside pressures (von Krogh et al., 2012), so the intrinsic-extrinsic contrast may cut across the distinction that matters. Second, incentives predict the quantity of performance better than its quality, while intrinsic motivation predicts quality better (Cerasoli et al., 2014), so an analysis of effort cannot detect an advantage in quality. Researchers should therefore use the autonomous-controlled contrast (Gagné & Deci, 2005) and study the quality of contributions alongside their amount.

## From Intention to Effort

The broadest pattern in the data is that motivation predicts stated intention better than actual effort, and that this gap is larger for intrinsic motives, especially altruism and community identification. Intentions often fail to become behavior when people lack the time, the opportunity, or a plan to act (Sheeran & Webb, 2016). Paid and career-related contribution tends to come with these resources, because an employer sets aside the hours, whereas volunteer contribution must compete with every other demand on the contributor's time. The comparison of measures points the same way: intrinsic motivation related less to effort recorded by platforms than to self-reported effort, while extrinsic motivation did not. Intrinsic motives may therefore be overstated in surveys, or strongly attached contributors may overstate their own effort. Because our gap test rests mainly on comparisons across samples, this conclusion should be confirmed within samples. Researchers should not use continuance intention as a stand-in for contribution, and they should measure contribution with digital trace data where possible.

## Reputation and Platform Design

Reputation related to effort on platforms with a reputation system but not on platforms without one. This is the clearest link in our data between a design feature and the strength of a motivation, and it fits signaling theory: a signal pays off only if others can observe it. A reputation system turns each contribution into visible points, ranks, or badges, so contributors who seek recognition can see the return on their work. The finding helps explain why the relationship between reputation and effort varies so widely across studies, and it is consistent with evidence that users work harder as they approach the next status level of a platform (Goes et al., 2016). The results also caution against a common shortcut. Platform reputation indicators, such as follower counts, varied far more across studies than the reputation motive and partly record past contribution, so studies that use them as measures of motivation risk mistaking an outcome for a cause. Because few platforms in our data lack a reputation system, this moderation should be confirmed, for example in field experiments that randomly show or hide contribution counts on user profiles.

## Monetary Rewards and Motivation Crowding

Monetary rewards were positively related to effort. The relationship was stronger for actual payment than for a desire for money and, in our exploratory comparison, stronger in OSS than in other settings. These patterns suggest that pay matters most when it buys time for contribution. In OSS, payment usually means a job that includes the contribution, whereas a prize in a contest rewards only its winners. We found no sign of motivation crowding, and two mechanisms may explain this. First, pay may sort rather than crowd. Performance pay attracts people who are good at the rewarded work and like it (Lazear, 2000), and OSS developers sort into projects that fit their motives (Belenzon & Schankerman, 2015). If firms hire the volunteers who already enjoy the work, pay and enjoyment will be positively correlated. Second, crowding theory itself predicts crowding in when rewards are seen as acknowledgment rather than control. Cross-sectional correlations cannot rule out crowding within individuals, so researchers should follow the same contributors before and after pay begins, for example when firms hire volunteer developers or when platforms introduce bounties.

## Reciprocity, Career, and Ability

Reciprocity and own use was the only antecedent not reliably related to effort. Users who contribute to fix their own problem often stop once the need is met (Shah, 2006), so the motive may predict a single contribution rather than sustained effort. The construct also joins a private need and an obligation to give back, which researchers should measure separately. Career and human capital had the weakest relationship with effort, which fits evidence that labor market returns in OSS come from recognized rank rather than from the volume of contribution (Hann et al., 2013). Self-efficacy and expertise related positively to both outcomes, as social cognitive theory predicts.

## Implications for Practice

Platform owners may expect reputation systems to channel reputation-seeking contributors into effort, but they should not read surveys of contributors' intentions as forecasts of contribution, especially when those intentions rest on enjoyment or belonging. Firms that sponsor OSS can pay developers without evidence that pay displaces their enjoyment, and the pay appears to work by giving them time for the work. Project maintainers can help identified and altruistic members act on their intentions by lowering the time and skill each contribution requires, for example with small, well-defined tasks and clear guidance for newcomers.

## Limitations and Future Research

First, our search was designed for OSS, so the studies of related settings form a comparison set rather than a complete sample, and our comparisons across settings are exploratory. The search was also limited to English, and IEEE Xplore returned no records, so studies published only there may be missing. Second, the relationships were highly heterogeneous, and differences in measurement add noise that weakens pooled correlations (Schmidt & Hunter, 2015). The gap test depends partly on one flagged value and was not significant in some subsets of studies, the trend in enjoyment depends on one study, and some subgroup estimates rest on few samples. Third, several relationships could not be pooled, including ideology (Stewart & Gosain, 2006), actual payment with continuance intention, and career and human capital on platforms without a reputation system, and we did not examine the quality of contributions or actual departures. Fourth, publication year is only a proxy for the professionalization of OSS, and moderators we did not code, such as firm sponsorship or project size, may also matter. Finally, most primary studies are cross-sectional surveys, so contribution may raise reputation, identification, and self-efficacy as much as these motives raise contribution. Studies that measure motivation before contribution and field experiments that vary rewards or recognition (Gallus, 2017) would offer stronger tests of cause.

# Conclusion

Motivation explains what contributors intend better than what they do, and this gap is widest for intrinsic motives. Intrinsic motivation was not more strongly related to effort than extrinsic motivation, money did not crowd out enjoyment, and reputation mattered mainly where platforms made contributions visible. The next step is to learn how platforms turn willingness into work.

# Declarations

**Data availability.** The search records, coded data, and R analysis script are available at Mendeley Data (DOI: 10.17632/kdkxpgtgr9.1). Appendices A to E follow the references.

**Use of generative AI.** During the preparation of this manuscript, the authors used Claude Opus 5.5 (Anthropic) to improve the grammar and the quality of the writing. The tool was not used to design the study, code the studies, analyze the data, or produce the results. The authors reviewed and edited all AI-assisted text and take full responsibility for the content of the manuscript.

**Conflict of interest.** The authors declare that they have no conflict of interest.

**Funding.** The authors received no funding for this work.

# Appendices

The appendices report the evidence behind the main text. Appendix A gives examples of excluded reports, Appendix B lists the included studies and their coding, and Appendix C reports the correlations each sample contributes. Appendix D reports the expanded analysis and the robustness checks, and Appendix E reports the moderator tests for each antecedent. The full coding sheet, which gives each coded statistic, its conversion, and its location in the report, and the analysis script are available at Mendeley Data (https://doi.org/10.17632/kdkxpgtgr9.1).

## Appendix A. Examples of Excluded Reports by Eligibility Criterion

Table A1 gives one or two examples of reports excluded under each of the eight eligibility criteria in the Method. Figure 2 reports how many reports were excluded under each criterion.

**Table A1. Examples of Excluded Reports**

| **Criterion** | **Example** | **Reason for exclusion** |
|---|---|---|
| 1. Original quantitative data | von Krogh et al. (2012) | A review of research on motivation in OSS development; it reports no original data |
| 2. Individual contributor as the unit of analysis | Stewart and Gosain (2006); Daniel et al. (2013) | Studied the effectiveness of development teams and the success of projects |
| 3. OSS or another open collaboration setting | Duan (2024) | Studied educators' adoption of open-source software, not contribution to a shared product |
| 4. At least one of the eight antecedents, measured as a characteristic of the contributor | Ho and Rai (2017); Schaarschmidt (2023) | Ho and Rai related project quality controls, not motivations, to continuance; Schaarschmidt's reputation measure rates the developer's employer, not the developer |
| 5. A relationship with contribution effort or continuance intention | Sojer and Henkel (2010); Bartsch et al. (2026) | The outcomes were code reuse and the intention to review one's own code |
| 6. The antecedent modeled as a predictor of contribution | Cai and Zhu (2016) | Developer reputation was the dependent variable |

| Criterion | Example | Reason for exclusion |
|---|---|---|
| 7. A zero-order correlation or a convertible statistic | Allyn and Misra (2009) | Reported only the model $R^2$ and which motives were significant |
| 8. Written in English | One AIS eLibrary report | AIS eLibrary records carry no language field, so the database filter did not remove it |

*Note:* References for these reports are in the main reference list or in the Appendix References.

## Appendix B. Included Studies and Samples

Table B1 lists the 41 included studies and their 46 independent samples, one sample per row. For studies with several reports of the same sample, the first report listed supplied the statistics, and the others are reports we merged with it. The Appendix References give the full reference for all 50 reports.

### Table B1. Included Studies and Their Samples

| Study | Setting | Year | Design | N | Antecedents | Outcomes | Analysis | Rewards | Reputation system | Status |
|---|---|---|---|---|---|---|---|---|---|---|
| Abhari et al. (2022) | Online community (open innovation) | 2022 | S | 264 | EN, CI, AL, RP, CH, MR | I | M, C2 | P | U | Published |
| Abu-Tayeh et al. (2018) | Civic reporting app | 2018 | S+R | 650 | AL, RO | Et | X | A | N | Published |
| Alexy and Leitner (2010) | OSS (scenario experiment) | 2010 | E | 201 | EN, MR | n/a | C2 | P | N | Grey |
| Badashian et al. (2014), GitHub | OSS (GitHub) | 2014 | R | 2,437,234 | PRI | Et | PRI | A | Y | Published |
| Badashian et al. (2014), Stack Overflow | Online community (Stack Overflow) | 2014 | R | 2,332,403 | PRI | Et | PRI | A | Y | Published |

| Study | Setting | Year | Design | N | Antecedents | Outcomes | Analysis | Rewards | Reputation system | Status |
|---|---|---|---|---|---|---|---|---|---|---|
| Balka et al. (2014) | Online community (open hardware) | 2014 | S | 309 | CI | Es | X | A | U | Published |
| Barcomb et al. (2019) | OSS (episodic volunteers) | 2019 | S | 101 | CI, CH | I | M | U | U | Published |
| Benbya and Belbaly (2010); also conference version | OSS (SourceForge) | 2010 | S | 122 | RO, RP, CH | Es | M | P | Y | Published |
| Bosu (2015) | OSS (code review) | 2015 | S | 287 | MR, SE | Es | M | P | U | Grey |
| Brown (2016) | OSS (Drupal) | 2016 | S | 131 | AL, RP | Es, I | M | U | Y | Grey |
| Choi and Yi (2015) | Online community (Yahoo! Answers) | 2015 | S+R | 150 | EN, AL, RO, RP, CH, SE | I | M | A | Y | Published |
| Choi et al. (2015) | OSS (Firefox users) | 2015 | S | 346 | CI | I | M | A | N | Published |
| Cox et al. (2018) | Citizen science (Zooniverse) | 2018 | S+R | 1,915 | AL, CH | Et | X | A | Y | Published |
| Daniel et al. (2018) | OSS | 2018 | S+R | 186 | CI | Et | M | P | U | Published |
| Daniel et al. (2020) | OSS | 2020 | S+R | 132 | CI, RO, CH | Et, I | M | U | Y | Published |
| Frey et al. (2011) | Crowdsourcing contest (open innovation) | 2011 | S+R | 104 | EN, MR | Et | M, C2 | P | U | Published |
| Hars and Ou (2002) | OSS | 2002 | S | 79 | EN, CI, AL, RO, RP, CH, MR | Es | M | P | N | Published |

| Study | Setting | Year | Design | N | Antecedents | Outcomes | Analysis | Rewards | Reputation system | Status |
|---|---|---|---|---|---|---|---|---|---|---|
| Hertel et al. (2003) | OSS (Linux kernel) | 2003 | S | 141 | EN, CI, RO, MR, SE | Es, I | M | P | N | Published |
| Ihl et al. (2019) | Crowdsourcing contest (idea contest) | 2019 | S+R | 470 | EN, RO, RP, MR | Et | M, C2 | P | N | Published |
| Ke and Zhang (2008); also Ke and Zhang (2009, 2010, 2011) | OSS | 2008 | S | 233 | EN, CI, AL, MR, SE | Es | M, C2 | P | U | Published |
| Khasraghi et al. (2017) | Crowdsourcing contest (Kaggle) | 2017 | S+R | 222 | EN | I | M | P | Y | Published |
| Lakhani (2006, ch. 2) | Crowdsourcing contest (InnoCentive) | 2006 | S | 295 | EN, RP, MR | Es | M, C2 | P | N | Grey |
| Lakhani and Wolf (2003); also Lakhani (2006, ch. 4) | OSS (SourceForge) | 2003 | S | 630 | EN, CI, RO, RP, CH, MR | Es | X | P | Y | Grey |
| Li et al. (2012); also Li et al. (2006) | OSS (SourceForge) | 2006 | S | 118 | EN, CI, MR | Es | M, C2 | U | U | Published |
| Luthiger and Jungwirth (2007) | OSS | 2007 | S | 1,330 | EN | Es, I | M | P | N | Published |
| Nov and Kuk (2008) | Wiki (experiment) | 2008 | E | 176 | EN, MR | n/a | C2 | A | N | Published |
| Nov et al. (2010) | Online community (Flickr) | 2010 | S+R | 276 | EN, CI, RP, CH | Et | X | A | Y | Published |
| Nov et al. (2014), BOINC | Citizen science | 2014 | S+R | 1,202 | EN, AL, RP, SE | Et | M | A | Y | Published |

| Study | Setting | Year | Design | N | Antecedents | Outcomes | Analysis | Rewards | Reputation system | Status |
|---|---|---|---|---|---|---|---|---|---|---|
| Nov et al. (2014), CWOP | Citizen science | 2014 | S+R | 1,837 | EN, AL, RP, SE | Et | M | A | Y | Published |
| Nov et al. (2014), Stardust@home | Citizen science | 2014 | S+R | 139 | EN, AL, RP, SE | Et | M | A | Y | Published |
| Racero et al. (2022) | OSS (teaching projects) | 2022 | S | 231 | EN, CI, CH, MR | I | X | P | N | Published |
| Restivo and van de Rijt (2014), group 1 | Wiki (Wikipedia) | 2014 | E | 200 | PRI | Et | PRI | A | Y | Published |
| Restivo and van de Rijt (2014), group 2 | Wiki (Wikipedia) | 2014 | E | 200 | PRI | Et | PRI | A | Y | Published |
| Restivo and van de Rijt (2014), group 3 | Wiki (Wikipedia) | 2014 | E | 200 | PRI | Et | PRI | A | Y | Published |
| Rivato and Zaggl (2023) | Online community (Stack Exchange) | 2023 | R | 8,678 | PRI | Et | PRI | A | Y | Published |
| Roberts (2006); also Roberts et al. (2006) | OSS (Apache) | 2006 | S+R | 288 | EN, RO, RP, MR | Et | M, C2 | P | Y | Grey |
| Santos et al. (2025) | OSS (students) | 2025 | S | 241 | EN, RP, CH | I | X | A | U | Published |
| Sun et al. (2012); also Sun et al. (2014) | Crowdsourcing contest (Taskcn) | 2012 | S | 205 | EN, MR, SE | I | M, C2 | P | U | Published |
| Taylor and Dantu (2022) | OSS (GitHub survey) | 2022 | S | 3,746 | CH, MR, SE | I | X | P | Y | Published |
| Wei (2026) | OSS (GitHub) | 2026 | S | 235 | EN, MR | n/a | C2 | P | Y | Grey |

| Study | Setting | Year | Design | N | Antecedents | Outcomes | Analysis | Rewards | Reputation system | Status |
|---|---|---|---|---|---|---|---|---|---|---|
| Wijnhoven et al. (2015) | Open government | 2015 | S | 161 | EN, AL, RO, RP, CH, MR, SE | I | M | A | N | Published |
| Wu (2007); also Wu et al. (2013) | OSS | 2007 | S | 214 | EN, CI, AL, RO, RP, CH, SE | Es, I | M | P | Y | Grey |
| Wu et al. (2007) | OSS | 2007 | S | 148 | AL, RO, CH | I | M | P | U | Published |
| Zhang, Liu, et al. (2026) | OSS | 2026 | S | 148 | AL, RP | I | M | A | U | Published |
| Zhang, Stol, et al. (2026); also Zhang, Qin, et al. (2024) | OSS (Rust) | 2024 | R | 272 | MR | Et | M | P | Y | Published |
| Zhou and Yuan (2023) | OSS | 2023 | S | 307 | EN, CI, RO, RP, SE | Es, I | M | A | Y | Published |

*Note:* Setting gives the platform or project where the study names it. Year = year of the earliest report of the sample. Design: S = survey; S+R = survey matched with platform records; R = platform records only; E = experiment. N = largest sample size reported for the sample; the two samples of Badashian et al. (2014) were capped at 10,000 in the analysis. Antecedents: EN = enjoyment-based intrinsic motivation; CI = community identification; AL = altruism; RO = reciprocity and own use; RP = reputation; CH = career and human capital; MR = monetary rewards; SE = self-efficacy and expertise; PRI = platform reputation indicator. Outcomes: Es = contribution effort, self-reported; Et = contribution effort from digital trace data; I = continuance intention. Antecedents and outcomes are those used in the main analysis, or in the expanded analysis for samples marked X. Analysis: M = main and expanded analyses; X = expanded analysis only; C2 = test of C2; PRI = robustness check on platform reputation indicators only. Rewards = monetary rewards in the setting: P = present; A = absent; U = unclear. Reputation system: Y = present; N = absent; U = unclear. Status: Published = journal article or conference paper; Grey = dissertation or working paper.

## Appendix C. Correlations by Sample

Tables C1 and C2 report the correlation that each sample contributes to the main analysis for each relationship. Where a sample reported several correlations for the same relationship, the value is their mean (see Method). Table C3 reports the partial correlations that the expanded analysis adds. Table C4 reports the correlations used to test C2, and Table C5 reports the platform reputation indicators used in the robustness check.

**Table C1. Correlations with Contribution Effort in the Main Analysis**

| Study | N | EN (H1a) | RO (H2a) | RP (H3a) | CH (H4a) | MR (H5a) | SE (H6a) | CI (H7a) | AL (H8a) |
|---|---|---|---|---|---|---|---|---|---|
| Benbya and Belbaly (2010) | 122 | | .15 | −.07 | .18 | | | | |
| Bosu (2015) | 287 | | | | | .19 | .14 | | |
| Brown (2016) | 124 | | | .39 | | | | | |
| Daniel et al. (2018) | 186 | | | | | | | .07 | |
| Daniel et al. (2020) | 132 | | −.04 | | .10 | | | −.11 | |
| Frey et al. (2011) | 104 | .18 | | | | .12 | | | |
| Hars and Ou (2002) | 79 | .07 | .30 | −.02 | .23 | .36 | | .12 | .19 |
| Hertel et al. (2003) | 141[a] | .01 | .16 | | | .54 | .30 | .32 | |
| Ihl et al. (2019) | 470 | .05 | .06 | .07 | | .11 | | | |
| Ke and Zhang (2008) | 204[b] | .20 | | | | .23 | .40 | .38 | .05 |
| Lakhani (2006, ch. 2) | 295 | .05 | | −.07 | | −.06 | | | |
| Li et al. (2012) | 118 | .30 | | | | .35 | | .30 | |
| Luthiger and Jungwirth (2007) | 1,330 | .12 | | | | | | | |
| Nov et al. (2014), BOINC | 1,202 | .41 | | .39 | | | .12 | | .20 |
| Nov et al. (2014), CWOP | 1,837 | .47 | | .58 | | | .01 | | .26 |

| Study | N | EN (H1a) | RO (H2a) | RP (H3a) | CH (H4a) | MR (H5a) | SE (H6a) | CI (H7a) | AL (H8a) |
|---|---|---|---|---|---|---|---|---|---|
| Nov et al. (2014), Stardust@home | 139 | .45 | | .14 | | | .17 | | .24 |
| Roberts (2006) | 288 | .11 | −.27 | .30 | | .22 | | | |
| Wu (2007) | 214[c] | .06 | .10 | .20 | .02 | | .03 | .26 | .22 |
| Zhang, Stol, et al. (2026) | 272 | | | | | .43 | | | |
| Zhou and Yuan (2023) | 307 | .68 | .69 | .67 | | | .70 | .72 | |

*Note:* Zero-order correlations. N = largest sample size among the sample's correlations with effort. Abbreviations as in Table B1. [a] N = 69 for monetary rewards and N = 39 for self-efficacy and expertise. [b] N = 233 for self-efficacy and expertise. [c] N = 181 for community identification.

**Table C2. Correlations with Continuance Intention in the Main Analysis**

| Study | N | EN (H1b) | RO (H2b) | RP (H3b) | CH (H4b) | MR (H5b) | SE (H6b) | CI (H7b) | AL (H8b) |
|---|---|---|---|---|---|---|---|---|---|
| Abhari et al. (2022) | 264 | .48 | | .39 | .43 | .38 | | .41 | .49 |
| Barcomb et al. (2019) | 101 | | | | .24 | | | .64 | |
| Brown (2016) | 131 | | | | | | | | .89 |
| Choi and Yi (2015) | 150 | .49 | .16 | .27 | .20 | | .36 | | .51 |
| Choi et al. (2015) | 346 | | | | | | | .62 | |
| Daniel et al. (2020) | 132 | | −.11 | | .05 | | | −.04 | |
| Hertel et al. (2003) | 141[a] | .19 | .40 | | | | .09 | .38 | |
| Khasraghi et al. (2017) | 222 | .23 | | | | | | | |

| Study | N | EN (H1b) | RO (H2b) | RP (H3b) | CH (H4b) | MR (H5b) | SE (H6b) | CI (H7b) | AL (H8b) |
|---|---|---|---|---|---|---|---|---|---|
| Luthiger and Jungwirth (2007) | 1,330 | .37 | | | | | | | |
| Sun et al. (2012) | 205 | .49 | | | | .36 | .30 | | |
| Wijnhoven et al. (2015) | 161 | .26 | .02 | .08 | −.01 | −.01 | .05 | | .07 |
| Wu (2007) | 214 | .38 | .19 | .25 | .19 | | | .78 | .34 |
| Wu et al. (2007) | 148 | | .30 | | .38 | | | | .42 |
| Zhang, Liu, et al. (2026) | 148 | | | .48 | | | | | .63 |
| Zhou and Yuan (2023) | 307 | .62 | .64 | .62 | | | .64 | .66 | |

*Note:* Zero-order correlations. N = largest sample size among the sample's correlations with intention. Abbreviations as in Table B1. The correlation of .89 in Brown (2016) is the flagged value dropped in one sensitivity check. [a] N = 39 for self-efficacy and expertise.

**Table C3. Partial Correlations Added in the Expanded Analysis**

| Study | Relationship | Partial correlation | N |
|---|---|---|---|
| Abu-Tayeh et al. (2018) | Reciprocity and own use → effort (H2a) | .11 | 650 |
| Abu-Tayeh et al. (2018) | Altruism → effort (H8a) | .09 | 650 |
| Balka et al. (2014) | Community identification → effort (H7a) | .26 | 309 |
| Cox et al. (2018) | Career and human capital → effort (H4a) | −.02 | 1,915 |
| Cox et al. (2018) | Altruism → effort (H8a) | .04 | 1,915 |
| Ke and Zhang (2008) | Reputation → effort (H3a) | .14 | 230 |
| Lakhani and Wolf (2003) | Enjoyment-based intrinsic motivation → effort (H1a) | .11 | 630 |

| Study | Relationship | Partial correlation | N |
|---|---|---|---|
| Lakhani and Wolf (2003) | Reciprocity and own use → effort (H2a) | .04 | 630 |
| Lakhani and Wolf (2003) | Reputation → effort (H3a) | .08 | 630 |
| Lakhani and Wolf (2003) | Career and human capital → effort (H4a) | .05 | 630 |
| Lakhani and Wolf (2003) | Monetary rewards → effort (H5a) | .13 | 630 |
| Lakhani and Wolf (2003) | Community identification → effort (H7a) | .07 | 630 |
| Nov et al. (2010) | Enjoyment-based intrinsic motivation → effort (H1a) | .10 | 276 |
| Nov et al. (2010) | Reputation → effort (H3a) | .11 | 276 |
| Nov et al. (2010) | Career and human capital → effort (H4a) | .13 | 276 |
| Nov et al. (2010) | Community identification → effort (H7a) | −.05 | 276 |
| Racero et al. (2022) | Enjoyment-based intrinsic motivation → intention (H1b) | .50 | 231 |
| Racero et al. (2022) | Career and human capital → intention (H4b) | .21 | 231 |
| Racero et al. (2022) | Monetary rewards → intention (H5b) | .18 | 231 |
| Racero et al. (2022) | Community identification → intention (H7b) | .14 | 231 |
| Santos et al. (2025) | Enjoyment-based intrinsic motivation → intention (H1b) | .22 | 241 |
| Santos et al. (2025) | Reputation → intention (H3b) | .13 | 241 |
| Santos et al. (2025) | Career and human capital → intention (H4b) | .09 | 241 |
| Taylor and Dantu (2022) | Career and human capital → intention (H4b) | .05 | 3,746 |
| Taylor and Dantu (2022) | Monetary rewards → intention (H5b) | −.03 | 3,746 |
| Taylor and Dantu (2022) | Self-efficacy and expertise → intention (H6b) | .10 | 3,746 |

*Note:* Partial correlations computed from regression results as $t/\sqrt{t^2 + df}$. The expanded analysis uses a partial correlation only when the sample reports no zero-order correlation for that relationship.

**Table C4. Correlations between Enjoyment-Based Intrinsic Motivation and Monetary Rewards (C2)**

| Study | N | r |
|---|---|---|
| Abhari et al. (2022) | 264 | .27 |
| Alexy and Leitner (2010) | 201 | .26 |
| Frey et al. (2011) | 104 | .02 |
| Ihl et al. (2019) | 470 | .40 |
| Ke and Zhang (2008) | 204 | .17 |
| Lakhani (2006, ch. 2) | 295 | −.21 |
| Li et al. (2012) | 118 | .01 |
| Nov and Kuk (2008) | 176 | .47 |
| Roberts (2006) | 288 | .04 |
| Sun et al. (2012) | 205 | .40 |
| Wei (2026) | 235 | .35 |

*Note:* Within-sample correlations. Pooled r = .21, 95% CI [.07, .34], k = 11, N = 2,560.

**Table C5. Platform Reputation Indicators and Contribution Effort**

| Study | Indicator | N | N used | r |
|---|---|---|---|---|
| Badashian et al. (2014), GitHub | Followers | 2,437,234 | 10,000 | .50 |
| Badashian et al. (2014), Stack Overflow | Followers | 2,332,403 | 10,000 | .91 |
| Restivo and van de Rijt (2014), group 1 | Barnstar award (randomized) | 200 | 200 | .23 |
| Restivo and van de Rijt (2014), group 2 | Barnstar award (randomized) | 200 | 200 | .07 |
| Restivo and van de Rijt (2014), group 3 | Barnstar award (randomized) | 200 | 200 | −.08 |
| Rivato and Zaggl (2023) | Accumulated score and platform credits | 8,678 | 8,678 | −.06 |

*Note:* N used = sample size after the cap of 10,000. In Restivo and van de Rijt (2014), the awards were assigned at random, so the indicator is an award received rather than a motive. Pooled r = .36, 95% CI [−.25, .76], k = 6.

## Appendix D. Expanded Analysis and Robustness Checks

Table D1 compares the main and expanded analyses for all 16 relationships. Only one conclusion changes: monetary rewards are no longer significantly related to continuance intention. Table D2 reports H9 and H10 under each alternative specification in the Method, and Table D3 reports both tests after dropping each sample in turn. H9 was not significant in any specification. The difference between the gaps in H10 stayed positive in every specification, but it was not significant in published studies only, in surveys only, with self-reported effort only, without the flagged correlation, or after dropping one of three samples. Table D4 compares measures of reputation and monetary rewards, and Table D5 repeats the moderator tests without the samples that most shaped them.

### Table D1. Main and Expanded Analyses

| Hypothesis | Antecedent | Outcome | Main k | Main r [95% CI] | Expanded k | Expanded N | Expanded r [95% CI] | p |
|---|---|---|---|---|---|---|---|---|
| H1a | Enjoyment-based intrinsic motivation | Effort | 14 | .24 [.12, .36] | 16 | 7,634 | .23 [.11, .33] | < .001 |
| H1b | Enjoyment-based intrinsic motivation | Intention | 9 | .40 [.31, .48] | 11 | 3,466 | .39 [.31, .47] | < .001 |
| H2a | Reciprocity and own use | Effort | 8 | .16 [−.10, .41] | 10 | 3,033 | .14 [−.04, .32] | .122 |
| H2b | Reciprocity and own use | Intention | 7 | .25 [.02, .45] | 7 | 1,253 | .25 [.02, .45] | .035 |
| H3a | Reputation | Effort | 11 | .26 [.09, .42] | 14 | 6,213 | .23 [.08, .37] | .003 |
| H3b | Reputation | Intention | 6 | .36 [.19, .52] | 7 | 1,485 | .33 [.16, .48] | < .001 |
| H4a | Career and human capital | Effort | 4 | .11 [.02, .20] | 7 | 3,368 | .07 [.00, .13] | .035 |
| H4b | Career and human capital | Intention | 7 | .22 [.09, .34] | 10 | 5,388 | .19 [.08, .28] | < .001 |
| H5a | Monetary rewards | Effort | 10 | .25 [.14, .35] | 11 | 2,816 | .23 [.14, .32] | < .001 |

| Hypothesis | Antecedent | Outcome | Main k | Main r [95% CI] | Expanded k | Expanded N | Expanded r [95% CI] | p |
|---|---|---|---|---|---|---|---|---|
| H5b | Monetary rewards | Intention | 3 | .25 [.01, .47] | 5 | 4,607 | .18 [−.02, .37] | .079 |
| H6a | Self-efficacy and expertise | Effort | 8 | .25 [.07, .42] | 8 | 4,258 | .25 [.07, .42] | .008 |
| H6b | Self-efficacy and expertise | Intention | 5 | .32 [.05, .54] | 6 | 4,608 | .28 [.05, .48] | .020 |
| H7a | Community identification | Effort | 8 | .28 [.04, .49] | 11 | 2,563 | .23 [.05, .40] | .012 |
| H7b | Community identification | Intention | 7 | .53 [.34, .68] | 8 | 1,736 | .49 [.29, .65] | < .001 |
| H8a | Altruism | Effort | 6 | .21 [.15, .26] | 8 | 6,240 | .16 [.08, .24] | < .001 |
| H8b | Altruism | Intention | 7 | .53 [.29, .70] | 7 | 1,216 | .53 [.29, .70] | < .001 |

*Note:* The expanded analysis adds partial correlations to the main analysis. p refers to the expanded estimate. k = number of independent samples; N = total sample size.

**Table D2. Tests of H9 and H10 under Alternative Specifications**

| Specification | Intrinsic r (k) | Extrinsic r (k) | H9 Δz | p | H10 Δz | p |
|---|---|---|---|---|---|---|
| Main analysis | .21 (16) | .28 (18) | −.07 | .369 | .28 | .033 |
| Expanded analysis (adds partial correlations) | .18 (21) | .25 (21) | −.07 | .386 | .30 | .017 |
| Self-reported effort only | .25 (8) | .28 (10) | −.03 | .811 | .24 | .170 |
| Effort from digital trace data only | .17 (8) | .28 (8) | −.11 | .287 | .32 | .031 |
| Published studies only | .23 (13) | .32 (13) | −.09 | .325 | .22 | .111 |
| Surveys only | .25 (8) | .28 (10) | −.03 | .811 | .26 | .140 |
| Reciprocity and own use added to the extrinsic group | .21 (16) | .26 (18) | −.05 | .540 | .26 | .047 |
| Without the flagged correlation of .89 (Brown 2016) | .21 (16) | .28 (18) | −.07 | .369 | .22 | .078 |
| Introjected regulation added to reputation (Li et al. 2012; Racero et al. 2022) | .21 (16) | .28 (18) | −.07 | .391 | .28 | .035 |

| Specification | Intrinsic r (k) | Extrinsic r (k) | H9 Δz | p | H10 Δz | p |
|---|---|---|---|---|---|---|
| Identified regulation scale added to career and human capital (Li et al. 2012) | .21 (16) | .28 (18) | −.08 | .359 | .28 | .032 |
| Without Zhou and Yuan (2023) | .17 (15) | .25 (17) | −.09 | .250 | .32 | .007 |
| Without the three citizen science samples (Nov et al. 2014) | .18 (13) | .26 (15) | −.08 | .383 | .29 | .038 |

*Note:* Intrinsic r and extrinsic r = pooled correlations of the two groups with contribution effort. Δz = difference in Fisher's z. For H9, it is the intrinsic group minus the extrinsic group. For H10, it is the intrinsic gap minus the extrinsic gap, where each gap is the group's Fisher's z with continuance intention minus its Fisher's z with contribution effort.

**Table D3. Tests of H9 and H10 with One Sample Dropped**

| Sample dropped | H9 Δz | p | H10 Δz | p |
|---|---|---|---|---|
| Abhari et al. (2022) | −.07 | .369 | .29 | .036 |
| Barcomb et al. (2019) | −.07 | .369 | .26 | .059 |
| Benbya and Belbaly (2010) | −.09 | .295 | .29 | .026 |
| Bosu (2015) | −.08 | .346 | .28 | .031 |
| Brown (2016) | −.07 | .427 | .21 | .091 |
| Choi and Yi (2015) | −.07 | .369 | .27 | .050 |
| Choi et al. (2015) | −.07 | .369 | .27 | .046 |
| Daniel et al. (2018) | −.06 | .442 | .27 | .041 |
| Daniel et al. (2020) | −.06 | .442 | .28 | .034 |
| Frey et al. (2011) | −.08 | .340 | .29 | .031 |
| Hars and Ou (2002) | −.07 | .382 | .28 | .035 |

| Sample dropped | H9 Δz | p | H10 Δz | p |
|---|---|---|---|---|
| Hertel et al. (2003) | −.05 | .521 | .28 | .040 |
| Ihl et al. (2019) | −.07 | .374 | .28 | .034 |
| Ke and Zhang (2008) | −.08 | .370 | .28 | .034 |
| Khasraghi et al. (2017) | −.07 | .369 | .30 | .023 |
| Lakhani (2006, ch. 2) | −.08 | .302 | .29 | .027 |
| Li et al. (2012) | −.08 | .378 | .28 | .035 |
| Luthiger and Jungwirth (2007) | −.07 | .429 | .28 | .042 |
| Nov et al. (2014), BOINC | −.07 | .433 | .28 | .044 |
| Nov et al. (2014), CWOP | −.06 | .431 | .27 | .038 |
| Nov et al. (2014), Stardust@home | −.09 | .282 | .30 | .025 |
| Roberts (2006) | −.07 | .432 | .27 | .041 |
| Sun et al. (2012) | −.07 | .369 | .29 | .040 |
| Wijnhoven et al. (2015) | −.07 | .369 | .27 | .039 |
| Wu (2007) | −.08 | .333 | .28 | .049 |
| Wu et al. (2007) | −.07 | .369 | .29 | .033 |
| Zhang, Liu, et al. (2026) | −.07 | .369 | .29 | .036 |
| Zhang, Stol, et al. (2026) | −.06 | .456 | .27 | .043 |
| Zhou and Yuan (2023) | −.09 | .250 | .32 | .007 |

*Note:* Main analysis, rerun after dropping the sample in the first column. Δz as in Table D2.

**Table D4. Measures of Reputation and Monetary Rewards**

| Measure | Outcome | Analysis | k | N | r [95% CI] |
|---|---|---|---|---|---|
| Reputation motive | Effort | Main | 11 | 5,077 | .26 [.09, .42] |
| Platform reputation indicators | Effort | Indicator samples | 6 | 29,278 | .36 [−.25, .76] |
| Actual payment | Effort | Main | 4 | 916 | .34 [.18, .48] |
| Actual payment | Effort | Expanded | 5 | 1,546 | .29 [.15, .42] |
| Actual payment | Intention | Main | 0 | n/a | n/a |
| Actual payment | Intention | Expanded | 1 | 3,746 | Not pooled |
| Expected or desired money | Effort | Main | 6 | 1,270 | .17 [.05, .30] |
| Expected or desired money | Effort | Expanded | 6 | 1,270 | .17 [.05, .30] |
| Expected or desired money | Intention | Main | 3 | 630 | .25 [.01, .47] |
| Expected or desired money | Intention | Expanded | 4 | 861 | .24 [.06, .40] |

*Note:* Actual payment = paid status, paid hours, or payment for OSS work. Expected or desired money = a stated expectation of, or desire for, a monetary return. Relationships reported by fewer than three samples were not pooled. The N for platform reputation indicators uses the cap of 10,000.

**Table D5. Moderator Tests without Influential Samples**

| Test | Samples | First group r (k) | Second group r (k) | Δz or slope | p |
|---|---|---|---|---|---|
| M3: reputation → effort, reputation system present vs. absent | All samples | .35 (8) | .00 (3) | .37 | < .001 |
| C1: enjoyment-based intrinsic motivation → effort, rewards present vs. absent | All samples | .10 (9) | .51 (4) | −.46 | < .001 |
| M3: reputation → effort, reputation system present vs. absent | Without Zhou and Yuan (2023) | .30 (7) | .00 (3) | .30 | .004 |

| Test | Samples | First group r (k) | Second group r (k) | Δz or slope | p |
|---|---|---|---|---|---|
| C1: enjoyment-based intrinsic motivation → effort, rewards present vs. absent | Without Zhou and Yuan (2023) | .10 (9) | .44 (3) | −.38 | < .001 |
| M3: reputation → effort, reputation system present vs. absent | Without Nov et al. (2014) | .33 (5) | .00 (3) | .33 | .035 |
| C1: enjoyment-based intrinsic motivation → effort, rewards present vs. absent | Without Nov et al. (2014) | .10 (9) | .68 (1) | Not testable | n/a |
| M2: enjoyment-based intrinsic motivation → effort, slope per decade | All samples | k = 14 | | .29 | .004 |
| M2: enjoyment-based intrinsic motivation → effort, slope per decade | Without Zhou and Yuan (2023) | k = 13 | | .18 | .078 |
| M2: enjoyment-based intrinsic motivation → effort, slope per decade | Without Nov et al. (2014) | k = 11 | | .24 | .024 |

*Note:* First group = reputation system present (M3) or monetary rewards present (C1); second group = absent. Δz = difference in Fisher's z between the groups. Slope = change in Fisher's z per decade of publication year, tested with the Knapp-Hartung adjustment. Not testable = fewer than two samples in a group. Zhou and Yuan (2023) reports the highest correlations, and Nov et al. (2014) supplies three citizen science samples.

## Appendix E. Moderator Results by Antecedent

Table E1 compares OSS with other settings for each antecedent and group (M1). Table E2 reports the publication year slopes (M2) for each antecedent. Both tables use the main analysis. The studies of other settings form a comparison set, so the setting comparisons are exploratory. The slopes other than those for monetary rewards and enjoyment-based intrinsic motivation with effort do not test a hypothesis and are also exploratory.

**Table E1. Setting Comparisons by Antecedent (M1)**

| Antecedent or group | Outcome | OSS r (k) | Other settings r (k) | Δz | p |
|---|---|---|---|---|---|
| Intrinsic motivation group | Effort | .20 (10) | .22 (6) | −.02 | .869 |
| Intrinsic motivation group | Intention | .54 (10) | .38 (5) | .21 | .104 |
| Extrinsic motivation group | Effort | .31 (12) | .23 (6) | .09 | .547 |
| Extrinsic motivation group | Intention | .35 (6) | .26 (4) | .09 | .507 |
| Enjoyment-based intrinsic motivation | Effort | .21 (8) | .28 (6) | −.07 | .590 |
| Enjoyment-based intrinsic motivation | Intention | .41 (4) | .40 (5) | .01 | .918 |
| Reciprocity and own use | Effort | .18 (7) | .06 (1) | Not testable | n/a |
| Reciprocity and own use | Intention | .31 (5) | .09 (2) | .23 | .172 |
| Reputation | Effort | .27 (6) | .24 (5) | .03 | .861 |
| Reputation | Intention | .46 (3) | .25 (3) | .24 | .178 |
| Career and human capital | Effort | .11 (4) | none (0) | Not testable | n/a |
| Career and human capital | Intention | .22 (4) | .22 (3) | .00 | .999 |
| Monetary rewards | Effort | .32 (7) | .05 (3) | .28 | < .001 |
| Monetary rewards | Intention | none (0) | .25 (3) | Not testable | n/a |
| Self-efficacy and expertise | Effort | .34 (5) | .08 (3) | .27 | .129 |
| Self-efficacy and expertise | Intention | .41 (2) | .24 (3) | .19 | .576 |
| Community identification | Effort | .28 (8) | none (0) | Not testable | n/a |
| Community identification | Intention | .55 (6) | .41 (1) | Not testable | n/a |
| Altruism | Effort | .15 (3) | .23 (3) | −.09 | .170 |
| Altruism | Intention | .63 (4) | .37 (3) | .35 | .211 |

*Note:* r (k) = pooled correlation and number of samples. Δz = Fisher's z in OSS minus Fisher's z in other settings. Not testable = fewer than two samples in a setting.

**Table E2. Publication Year Slopes by Antecedent (M2)**

| **Antecedent** | **Outcome** | **k** | **Years** | **Slope per decade** | **p** | **Residual τ²** |
|---|---|---|---|---|---|---|
| Enjoyment-based intrinsic motivation | Effort | 14 | 2002-2023 | .29 | .004 | .031 |
| Enjoyment-based intrinsic motivation | Intention | 9 | 2003-2023 | .15 | .097 | .016 |
| Reciprocity and own use | Effort | 8 | 2002-2023 | .15 | .375 | .123 |
| Reciprocity and own use | Intention | 7 | 2003-2023 | .01 | .953 | .109 |
| Reputation | Effort | 11 | 2002-2023 | .29 | .040 | .072 |
| Reputation | Intention | 6 | 2007-2026 | .23 | .120 | .031 |
| Career and human capital | Effort | 4 | 2002-2020 | −.03 | .781 | .005 |
| Career and human capital | Intention | 7 | 2007-2022 | −.01 | .969 | .034 |
| Monetary rewards | Effort | 10 | 2002-2024 | −.01 | .903 | .031 |
| Monetary rewards | Intention | 3 | 2012-2022 | Not estimated (k < 4) | n/a | n/a |
| Self-efficacy and expertise | Effort | 8 | 2003-2023 | .23 | .223 | .053 |
| Self-efficacy and expertise | Intention | 5 | 2003-2023 | .32 | .128 | .038 |
| Community identification | Effort | 8 | 2002-2023 | .07 | .650 | .125 |
| Community identification | Intention | 7 | 2003-2023 | −.09 | .663 | .109 |
| Altruism | Effort | 6 | 2002-2014 | .09 | .323 | .002 |
| Altruism | Intention | 7 | 2007-2026 | .17 | .538 | .159 |

*Note:* Slope = change in Fisher's z per decade of publication year, centered at 2015, from a random-effects meta-regression tested with the Knapp-Hartung adjustment. Residual τ² = between-study variance that publication year does not explain. We ran the meta-regression only when at least four samples were available.

## Appendix References

This list gives every report of an included study, including those the main text also cites, and the excluded reports in Appendix A that the main reference list does not contain.